\documentclass[pra,showpacs]{revtex4}
 \usepackage{amssymb} \usepackage{graphicx}

\begin{document}
\title{Basic Gravitational Currents and Killing-Yano Forms}
\author{\"{O}. A\c{c}{\i}k
$^{1}$}\email{ozacik@science.ankara.edu.tr}
\author{\"{U}. Ertem $^{1}$}
\email{uertem@science.ankara.edu.tr}
\author{M. \"{O}nder $^{2}$}\email{onder@hacettepe.edu.tr}
 \author{A. Ver\c{c}in $^{1}$}
 \email{vercin@science.ankara.edu.tr}
%\thanks{E.mail:vercin@science.ankara.edu.tr}%
\address{$^{1}$ Department of Physics, Ankara University, Faculty of Sciences,
06100, Tando\u gan-Ankara, Turkey\\
$^{2}$ Department of Physics Engineering, Hacettepe University,
06532, Beytepe-Ankara, Turkey.}

\date{\today}

\begin{abstract}

It has been shown that for each Killing-Yano (KY)-form accepted by
an $n$-dimensional (pseudo)Riemannian manifold of arbitrary
signature, two basic gravitational currents can be defined.
Conservation of the currents are explicitly proved by showing
co-exactness of the one and co-closedness of the other. Some general
geometrical facts implied by these conservation laws are also
elucidated. In particular, the conservation of the one-form currents
implies that the scalar curvature of the manifold is a flow
invariant for all of its Killing vector fields. It also directly
follows that, while all KY-forms and their Hodge duals on a constant
curvature manifold are the eigenforms of the Laplace-Beltrami
operator, for an Einstein manifold this is certain only for KY
$1$-forms, $(n-1)$-forms and
their Hodge duals.\\


\end{abstract}


\maketitle

\section{INTRODUCTION}

Conservation laws are intimately related to the symmetries of the
underlying space-time. Killing vector fields which generate local
isometries play a prominent role in constructing the usual conserved
currents and a conserved charge is associated with each such a
current. In an $n$-dimensional space-time, by making use of the
asymptotic symmetries, an ADM conserved charge \cite{Adm} can be
written as an integral over $(n-2)$-spheres at spatial infinity
\cite{Abbott-Deser}. For example, time translation Killing vector
field defines the ADM mass and rotational Killing vector fields
define the ADM angular momenta. Antisymmetric generalizations of
Killing vector fields to higher order forms are called Killing-Yano
(KY) forms. These reflect the hidden symmetries of the metric and
can be used in generalizing the conserved currents to $p$-brane
space-times that are extended objects with $p$ spatial dimensions
\cite{Kastor}. For the most recent and for the earlier seminal
references about KY (and conformal KY) forms we refer to three
recent PhD dissertations \cite{Kress,Semmelman2,Kubiznak}.

Abbot-Deser (AD) construction of conserved charges
\cite{Abbott-Deser} can be extended to KY-forms and in such a case
the associated generalized charges are called Y-ADM charges. These
are constructed for asymptotically flat and asymptotically anti-de
Sitter space-times \cite{Kastor,Cebeci}. In ADM case, the charges
can be written in terms of Einstein forms and therefore they can be
related to stress-energy forms via Einstein equations. The physical
meanings of these charges are obtained from this relation. On the
other hand, for the generalized charges, there is no direct relation
with Einstein equations and hence their physical interpretations are
not yet clear. The ADM charges are extensive quantities since they
are obtained from integrals over the $(n-2)$-spheres at infinity
which enclose $(n-1)$-dimensional regions. However, a Y-ADM charge
associated to a KY $p$-form is calculated from the integrals over
$(n-p-2)$-spheres at spatial infinity along the directions
transverse to the $p$-brane. Since these spheres constitute the
boundaries of $(n-p-1)$-dimensional regions, the corresponding
charges are intensive quantities. This makes it possible to
interpret Y-ADM charges as charge densities for $p$-brane
space-times.

A conventional way of defining a conserved current and the
corresponding conserved charge in an $n$-dimensional space-time can
be described as follows. Suppose that a $p$-form $J$ satisfies
$d^{\ast}J=0$, where $d$ stands for the exterior derivative and
$^{\ast}$ denotes the Hodge map. Then, by the Stokes theorem the
integral of $^{\ast}J$ over the boundary of a $(n-p+1)$-dimensional
region vanishes. This means that $J$ is a conserved current and by
integrating $^{\ast}J$ over an arbitrary $(n-p)$-dimensional region
$\Sigma$, we can define the corresponding charge by
$Q=\int_{\Sigma}{}^{\ast}J$. $Q$ is a conserved charge in the sense
that it is a constant for all regions sharing the same homology
class with $\Sigma$. A $p$-form $J$  is co-closed if and only if
$\delta J=0$, where $\delta$ represents the co-derivative operator
defined below. Noting that the co-closedness of $J$ is equivalent to
the condition $d^{\ast}J=0$, the above remarks can be generalized as
follows: any co-closed $p$-form is a conserved current and the
corresponding conserved charge is defined as prescribed above.

Evidently, all linear combinations of conserved quantities are
trivially conserved and therefore a conservation law must be related
to a basic current that can not be constructed from others.
Physically meaningful conserved charges must also be constructed
from the basic currents. In two recent papers \cite{Kastor,Cebeci},
it has been shown that the $p$-form
\begin{eqnarray}
{\cal J}=-i_{X_{a}}i_{X_{b}}\omega\wedge R^{ab}+2(-1)^p
i_{X_{a}}\omega\wedge P^{a}+\cal{R}\omega\;,
\end{eqnarray}
constructed from a KY $p$-form $\omega$ (for definition see equation
(5)) and curvature characteristics is a generalized conserved
current and can be related to a conserved charge of a $p$-brane
space-time. Throughout this study we assume the Einstein summation
convention over repeated indices and we use $i_{X}$ to denote the
interior derivative with respect to the vector field $X$. In the
expression (1), $R^{ab}$ are the curvature 2-forms and
\begin{eqnarray}
P^{a}=i_{X_{b}} R^{ba}\;,\quad {\cal R}=i_{X_{a}}P^{a}\;,
\end{eqnarray}
denotes, respectively, the Ricci 1-forms and the curvature scalar.
The current ${\cal J}$ is an attempt to generalize the well-known
($1$-form) current $K_{a}{}^{\ast^{-1}}G^{a}$ which is constructed
from the components of a Killing vector field $K$ and Einstein
$(n-1)$-forms $G^{a}=R_{bc}\wedge^{\ast}(e^{abc})$. Here
$^{\ast^{-1}}$ denotes the inverse of the Hodge map which is defined
as $s(-1)^{p(n-p)}$ ($s$ is the sign of the determinant of metric)
times the Hodge map when acting on $p$-forms.

In this study we will show that two fundamental currents can be
constructed from KY-forms and curvature characteristics of the
underlying space-time and prove that ${\cal J}$ given by (1) is a
particular linear combination of gravitational currents proposed
here. We furthermore show that one of the currents is co-exact which
enables us to discover a number of basic geometric facts some of
which seem to be unknown in the literature. These facts can be
stated as follows. (i) The scalar curvature of any
(pseudo)Riemannian manifold is a flow invariant for all of its
Killing vector fields. (ii) The conserved currents directly provide
decompositions for differential forms constructed from
``contractions" of KY $p$-forms with the Ricci $1$-forms, curvature
$2$-forms and with the Einstein $(n-1)$-forms (for the values
$p=1,2$ and $p=n-1$). These contractions are expressible solely in
terms of KY-forms themselves and their (co)derivatives. (iii) On
Einstein manifolds, duals of all Killing and Yano vector fields are
eigenforms of the Laplace-Beltrami operator such that the
eigenvalues have multiplicities with well defined lower bounds. Some
well known facts related to the spectrum of Laplace-Beltrami
operator on constant curvature manifolds also follow directly from
the properties of conserved currents.

From here on we assume that the underlying manifold is an
$n$-dimensional (pseudo)Riemannian manifold with arbitrary
signature. Some of our conventions and notations are fixed above for
which, as well as for the remaining ones, we have mainly adopted the
conventions of reference \cite{Benn-Tucker}. The covariant
derivative with respect to a given vector field $X$ will be denoted
by $\nabla_{X}$ in terms of which the exterior derivative and
co-derivative can be expressed as
\begin{eqnarray}
d=e^{a}\wedge\nabla_{X_a}\;,\quad \delta=-i_{X^{a}}\nabla_{X_a}\;,
\end{eqnarray}
where the local co-frame $\{e^{a}\}$ is dual to the tangent frame
$\{X_{a}\}$ such that $i_{X_{b}}e^{a}=e^{a}(X_b)=\delta^{a}_{b}$.
The action of  $i_{X}$ on an arbitrary $p$-form $\alpha$ is defined,
for all vector fields $Y_{j}$, by
$(i_{X}\alpha)(Y_1,\dots,Y_{p-1})=p \alpha(X,Y_1,\dots,Y_{p-1})$.
When acting on an arbitrary $p$-form $\delta$ can be written as
$(-1)^{p}\;^{\ast^{-1}}d^{\ast}$.

The rest of the paper is organized as follows. The definitions of
two basic (uncomposite) currents and the proofs of co-exactness of
the one and co-closedness of the other are given in the next
section. Three main operators that play important role in our study
are also introduced there and their relevant properties are
presented. The basic $p$-form currents for the values
$p=1,\;n-2,\;n-1$, are studied in Section III where some of the
general geometric facts, such as the statements (i) and (ii) given
above, are explored as well. Conserved currents on some special
manifolds, such as Ricci flat, conformally flat and constant
curvature manifolds as well as Einstein manifolds, are considered in
Section IV. A brief summary of the study and some concluding remarks
are given in the last section. Contracted Bianchi identities that
play prominent role in the proof of the co-closedness of a basic
current are collected in Appendix A and an alternative proof for the
statement (i) given above is provided in Appendix B.

\section{Two Basic Conserved Currents}

We begin by defining three $p$-forms
\begin{eqnarray}
j_{1}&=&i_{X_{a}}i_{X_{b}}\omega\wedge R^{ab}\;,\nonumber\\
j_{2}&=&i_{X_{a}}\omega\wedge P^{a}\;,\\
j_{3}&=&\cal{R}\omega\;,\nonumber
\end{eqnarray}
for a given KY $p$-form $\omega$. A $p$-form $\omega$ is called a KY
$p$-form if and only if
\begin{eqnarray}
\nabla_{X}\omega=\frac{1}{p+1}i_{X}d\omega\;,
\end{eqnarray}
is satisfied for all vector fields $X$. Two immediate consequences
of this definition are that, all KY-forms are co-closed and satisfy
the relation $i_{Y}\nabla_{X}\omega+i_{X}\nabla_{Y}\omega=0$, for
any two vector fields $X$ and $Y$. The second relation means that
all symmetrized covariant derivatives of KY-forms vanish. As is
evident from the definition (5), any function is a KY $0$-form and a
KY $n$-form is a constant multiple of the volume form. As
manifestations of the hidden symmetries of the metric of the
underlying space-time, co-closedness of KY-forms means that all of
them define conserved currents and corresponding conserved charges.
Apart from this well-established fact, as we are about to see, their
particular combinations with the curvature characteristics, also
give rise to important conservation laws.

The main goal of our study is to prove that the ($p$-form) currents
\begin{eqnarray}
{\cal J}_{1}&=&-j_1+(-1)^pj_2\;,\\
{\cal J}_{2}&=&(-1)^pj_2+j_3\;,
\end{eqnarray}
are separately conserved on any (pseudo)Riemannian manifold for all
$p$'s. These are the currents that were, and will be, referred to as
basic. We shall also prove that ${\cal J}_{1}$ is co-exact and
elucidate its important geometrical and physical implications.
Evidently all linear combinations of ${\cal J}_{1}$ and ${\cal
J}_{2}$ are also conserved and the conservation of (1) is a
particular case of this fact since ${\cal J}={\cal J}_{1}+{\cal
J}_{2}$. These enable us to define more conserved charges and of
course physically meaningful and fundamental ones must be directly
constructed from ${\cal J}_{1}$ and ${\cal J}_{2}$. Finally, we
should note that (6) and (7) can be rewritten more compactly as
\begin{eqnarray}
{\cal J}_{1}=i_{X_{a}}(i_{X_{b}}\omega\wedge R^{ba})\;,\quad {\cal
J}_{2}= (-1)^pi_{X_a}(\omega\wedge P^{a})\;.
\end{eqnarray}
These clearly exhibit that for all KY $0$-forms (for all functions)
while ${\cal J}_{1}$ is identically zero, ${\cal J}_{2}$ is equal to
$j_3$ and that both currents are identically zero for KY $n$-forms.
Moreover, by observing that the current (1) is
\begin{eqnarray}
{\cal J}=i_{X_{a}}i_{X_{b}}(\omega\wedge R^{ba})\;,\nonumber
\end{eqnarray}
we see that for all KY $(n-1)$-forms ${\cal J}$ is identically zero.
But, in our case the currents are non-vanishing but linearly
dependent. So, for a general (pseudo)Riemannian manifold we can
safely say that, the space of conserved currents that linearly
depend on a given KY p-form is, for $0<p<n-1$, at least two
dimensional and is spanned by ${\cal J}_{1}$ and ${\cal J}_{2}$. Of
course, in some special cases these may be linearly dependent for
some, or even for all values of $p$. For example, as we have shown
in Section IV, on the constant curvature manifolds the latter
extreme case is observed. More details in this context are given in
sections III and IV.

\subsection{Main Operators}

The following three second order differential operators
\begin{eqnarray}
\mathbf{R}(X_a,X_b)&=&\nabla_{X_a}\nabla_{X_b}-\nabla_{X_b}\nabla_{X_a}-\nabla_{[X_a,X_b]}\;,\\
I(\textbf{R})&=&e^{a}\wedge i_{X_b}\mathbf{R}(X^{b},X_a)\;,\\
\nabla^2(X_a,X_b)&=&\nabla_{X_a}\nabla_{X_b}-\nabla_{\nabla_{X_a}X_b}\;,
\end{eqnarray}
play prominent roles in this study. The first is the well-known
curvature operator of the geometry, the second is known as the
curvature endomorphism \cite{Semmelman1} and the third is the
Hessian. All of these operators are degree-preserving when acting on
differential forms. The torsion-zero condition
\begin{eqnarray}
[X,Y]=\nabla_{X}Y-\nabla_{Y}X\;,
\end{eqnarray}
of (pseudo)Riemannian geometry implies that the curvature operator
can be written as the anti-symmetric difference
\begin{eqnarray}
\mathbf{R}(X_a,X_b)=\nabla^2(X_a,X_b)-\nabla^2(X_b,X_a)\;,
\end{eqnarray}
of two Hessians. The most important property of the curvature
endomorphism is that it can also be written as the difference
$I(\textbf{R})=\nabla^2-\displaystyle{\not}d^2$ (the classical
Weitzenb\"{o}ck formula) of the trace of Hessian
$\nabla^2=\nabla^2(X_a,X^{a})$ and the well-known Laplace-Beltrami
operator
\begin{eqnarray}
\displaystyle{\not}d^{2}=-(\delta d+d\delta) \;.\nonumber
\end{eqnarray}

An important property of the Hessian that will be used in our
analysis is that its double contraction vanishes
 \begin{eqnarray}
i_{X_b}i_{X_a}\nabla^2(X^{b},X^{a})=0.
\end{eqnarray}
This can be easily proved by using the definition of co-derivative
and the general relation
\begin{eqnarray}
[\delta, i_{X}]_{+}=-i_{X^{a}}i_{\nabla_{X_{a}}X}\;,
\end{eqnarray}
where $[,]_{+}$ denotes the anti-commutator. This relation easily
follows from the definition of $\delta$ and the relation $[\nabla_X
, i_Y]=i_{\nabla_XY}$. Then by direct computation we obtain
\begin{eqnarray}
i_{X_a}i_{X_b}\nabla_{X^{b}}\nabla_{X^{a}}&=&-i_{X_a}\delta\nabla_{X^{a}}
=-{\omega^{c}}_{a}(X_b)i_{X^b}i_{X_c}\nabla_{X^{a}}\nonumber\\
&=&i_{X_{a}}i_{X_b}\nabla_{\nabla_{X^{b}}X^{a}}\;,\nonumber
\end{eqnarray}
which proves (14). Here we have used the Poincare lemma $\delta^2=0$
and $\omega^{c}_{\;a}$'s are the connection $1$-forms defined by
$\nabla_{X_{b}}X_{a}=\omega^{c}_{\;a}(X_b)X_{c}$. As an aside, from
(13) we see that the double contraction of the curvature operator
also vanishes.

\subsection{Co-exactness of ${\cal J}_{1}$}

We will show that the co-exactness of ${\cal J}_{1}$ is a direct
result of the action of curvature endomorphism on KY-forms and the
co-closedness of ${\cal J}_{2}$ also follows from action of these
operators on KY-forms and Bianchi identities. The actions of
curvature operator and curvature endomorphism on an arbitrary
$p$-form $\phi$ are as follows
\begin{eqnarray}
\mathbf{R}(X_a,X_b)\phi&=&-i_{X_{c}}R_{ab}\wedge i_{X^{c}}\phi\;,\\
I(\textbf{R})\phi&=&P_{c}\wedge i_{X^{c}}\phi-R_{cb}\wedge
i_{X^{b}}i_{X^{c}}\phi\;.
\end{eqnarray}
The first relation can easily be found in the literature (see
\cite{Benn-Tucker} equation (8.1.11) and \cite{Benn-Kress}) and the
second follows from the first. Indeed, multiplying (16) by
$e^{a}\wedge$ and using the first Bianchi identity $R_{ab}\wedge
e^{b} =0$ we find
\begin{eqnarray}
e^{a}\wedge \mathbf{R}(X_a,X_b)\phi=-R_{cb}\wedge i_{X^{c}}\phi\;,
\end{eqnarray}
and then by contracting with $i_{X^b}$ we arrive at (17). A
comparison of the right hand side of (17) with (6) shows that ${\cal
J}_{1}$ is generated by the action of the curvature endomorphism on
the KY $p$-form $\omega$ used in its definition:
\begin{eqnarray}
{\cal J}_{1}=-I(\textbf{R})\omega\;.
\end{eqnarray}

The last relation implies that the action of other operators on
KY-forms may also have important implications. It is convenient to
consider the action of Hessian first. By differentiating both sides
of the defining relation (5) we obtain
\begin{eqnarray}
\nabla^2(X_a,X_b)\omega&=&\frac{1}{p+1}i_{X_b}\nabla_{X_a}d\omega\;,\\
\mathbf{R}(X_a,X_b)\omega&=&\frac{1}{p+1}(i_{X_b}\nabla_{X_a}-i_{X_a}\nabla_{X_b})d\omega\;.
\end{eqnarray}
The second relation is obtained from (20) by virtue of (13).
Multiplying the both sides of (21) with $e^{a}\wedge$, leads us to
\begin{eqnarray}
e^{a}\wedge \mathbf{R}(X_a,X_b)\omega
=-\frac{p}{p+1}\nabla_{X_b}d\omega\;,
\end{eqnarray}
where we have used the scaling property $e^{a}\wedge
i_{X_a}\alpha=k\alpha$ that holds for any $k$-form $\alpha$. By
contracting both sides of (22) with $i_{X^b}$, we obtain
\begin{eqnarray}
I(\textbf{R})\omega=\frac{p}{p+1}\delta d\omega\;,
\end{eqnarray}
and by comparing with (19) we arrive at
\begin{eqnarray}
{\cal J}_{1}=-\frac{p}{p+1}\delta d\omega\;.
\end{eqnarray}
That is, ${\cal J}_{1}$ is a co-exact $p$-form and hence provides a
conserved current.

\subsection{Co-Closedness of ${\cal J}_{2}$}

In proving the co-closedness of ${\cal J}_{2}$ we shall need the
covariant derivatives of $d\omega$ and the contracted Bianchi
identities. Some of these identities (that can be derived from the
so-called second Bianchi identity) are not easily found in the
literature and as they are repeatedly used below, we have collected
them in the Appendix A where their derivations are given in some
details.

From (18) and (22) we directly read
\begin{eqnarray}
\nabla_{X_b}d\omega=\frac{p+1}{p}R_{cb}\wedge i_{X^{c}}\omega\;,
\end{eqnarray}
and by taking the covariant derivatives of both sides we obtain
\begin{eqnarray}
\nabla_{X_b}\nabla_{X_a}d\omega&=&\frac{p+1}{p}(\nabla_{X_b}R^{c}{}_a\wedge
i_{X_c}\omega+R^{c}{}_a\wedge \nabla_{X_b}i_{X_c}\omega)\;.
\end{eqnarray}
We now define
\begin{eqnarray}
T^{bca}&=&\nabla_{X^b}R^{ca}+\omega^{cl}(X^b){R_{l}}^{a}+\omega^{al}(X^{b}){R^{c}}_{l}\;,\\
Q^{ba}&=&T^{bca}\wedge i_{X_c}\omega\;.
\end{eqnarray}
$T^{bca}$ is thoroughly investigated in Appendix A. In terms of
$Q^{ba}$ we rewrite (26) as
\begin{eqnarray}
\nabla_{X^b}\nabla_{X^a}d\omega=\frac{p+1}{p}Q^{ba}-
\omega^{al}(X^b)\nabla_{X_l}d\omega+\frac{1}{p}R^{ca}\wedge
i_{X_c}i_{X^b}d\omega\nonumber\;,
\end{eqnarray}
and by leaving $Q^{ba}$ alone we arrive at
\begin{eqnarray}
Q^{ba}=\frac{1}{p+1}[p\nabla^2(X^b,X^a)-R^{ca}\wedge
i_{X_c}i_{X^b}]d\omega\;.
\end{eqnarray}

We are now ready to take the co-derivative of ${\cal J}_{2}$ given
by (8)
\begin{eqnarray}
\delta {\cal J}_{2}=-(-1)^pi_{X_b}\nabla_{X^{b}}i_{X_a}(\omega\wedge
P^{a})\nonumber.
\end{eqnarray}
By making use of the relation $[\nabla_X , i_Y]=i_{\nabla_XY}$, the
definition of the KY-forms and contracted Bianchi identity
$i_{X_b}P^{a}=i_{X_a}P^{b}$, we can write
\begin{eqnarray}
\delta {\cal J}_{2}&=&-(-1)^pi_{X_a}i_{X_b}(\omega\wedge
S^{ab})\;,\nonumber
\end{eqnarray}
where we have defined
$S_{ab}=\nabla_{X_{a}}P_{b}+\omega_{bk}(X_a)P^{k}$. In Appendix A,
it is shown that $S^{ab}=S^{ba}+i_{X_c}T^{cab}$ which when
substituted into above relation yields
\begin{eqnarray}
\delta {\cal J}_{2}=-\delta {\cal
J}_{2}-(-1)^pi_{X_a}i_{X_b}[\omega\wedge i_{X_c}T^{cab}]\;.
\end{eqnarray}
Thus, by making use of the cyclic property of $T^{abc}$ given by
(A12), we have
\begin{eqnarray}
2\delta {\cal J}_{2}&=&-i_{X_a}i_{X_b}[i_{X_c}(\omega\wedge
T^{cab})-i_{X_c}\omega\wedge T^{cab}]\nonumber\\
&=&i_{X_a}i_{X_b}(i_{X_c}\omega\wedge T^{cab})\nonumber\\
&=&-i_{X_a}i_{X_b}[i_{X_c}\omega\wedge
(T^{abc}+T^{bca})]\nonumber\;,
\end{eqnarray}
and the anti-symmetry property of $T^{cab}$ with respect to the last
two indices (see (A7)), we end up with
\begin{eqnarray}
\delta {\cal J}_{2}=-i_{X_a}i_{X_b}Q^{ba}\;.
\end{eqnarray}

So, the proof of the co-closedness of ${\cal J}_{2}$ has been
reduced to the vanishing of double contraction of $Q^{ba}$. We now
observe that, by virtue of (14), the double contraction of the first
term at the right hand side of (29) vanishes. For the double
contraction of the second term of (29) we obtain
\begin{eqnarray}
i_{X_a}i_{X_b}(R^{ca}\wedge i_{X_c}i_{X^b}d\omega)
&=&i_{X_b}P^{c}\wedge i_{X_c}i_{X^{b}}d\omega+i_{X_b}R^{ca}\wedge
i_{X_a}i_{X_c}i_{X^{b}}d\omega\nonumber\;.
\end{eqnarray}
Since $i_{X_b}P_{c}$ is symmetric, the first term at the right hand
side vanishes and as the cyclic sum of $i_{X^b}R^{ca}$ amounts to
zero (see (A4)), the second term vanishes as well. These prove
$i_{X_a}i_{X_b}Q^{ab}=0$ and hence ${\cal J}_{2}$ is co-closed.

\section{Special $p$-form Currents: General Geometric Facts}

In this section we study the basic $p$-form currents for the values
$p=1,\;n-2,\;n-1$, and explore some of the general implications that
follow from their conservations. Some of these implications which
hold for any (pseudo) Riemannian manifold were concisely stated in
the introduction by the statements (i) and (ii). It is worth
emphasizing that the special cases considered below exhaust all
possible cases in four dimensions. Similar investigations for some
physically  relevant special manifolds are carried out in the next
section.

\subsection{One-form Currents}

Evidently, for all KY $1$-forms $j_1$ is identically zero and
therefore, $j_2$ is co-exact. For a given KY $1$-form $\omega$ we
have $j_2=K_aP^{a}$, where
\begin{eqnarray}
K=(i_{X_a}\omega)X^{a}=K^{a}X_a \;,\nonumber
\end{eqnarray}
is the Killing vector field dual to $\omega$, and then from (6) and
(24) we obtain
\begin{eqnarray}
K_aP^{a}=\frac{1}{2}\delta d\tilde{K}\;.
\end{eqnarray}
This relation is already known in the literature (see pp.231 in
\cite{Benn-Tucker}) and the above result may be considered as an
alternative derivation. What is more important is that the
co-closedness of ${\cal J}_{1}$ is equivalent to $\delta j_2=0$ and
this implies that $j_3$ must also be co-closed, for ${\cal J}_{2}$
is co-closed. On the other hand, from the definition
$j_3=\omega{\cal R}$ we find
\begin{eqnarray}
\delta j_3=-i_{X_a}\nabla_{X^{a}}(\omega{\cal R})=(\delta\omega)
{\cal R}-i_{X_a}\omega\nabla_{X^{a}}{\cal R}\nonumber
\end{eqnarray}
and since all KY-forms are co-closed, we arrive at $\delta
j_3=-\nabla_{K}{\cal R}=0$. As a result, the covariant derivative of
curvature scalar with respect to any Killing vector field $K$ of the
underlying manifold must be zero;
\begin{eqnarray}
\nabla_{K}{\cal R}=0\;.
\end{eqnarray}
So we have arrived at a general relation of the (pseudo)Riemannian
geometry which uncovers an important property of the Killing vector
fields that seems to be, as far as we know, unnoticed in the
literature. Since the Lie derivative and covariant derivative
coincide on $0$-forms, this also means that the Lie derivative of
the scalar curvature with respect to Killing vector fields is zero
for any (finite) dimension and signature. More precisely, the scalar
curvature of any (pseudo)Riemannian manifold is a flow-invariant for
all of its Killing vector fields. An alternative proof of equation
(33) is given in Appendix B.

It is worth mentioning that for $1$-form currents above we have
\begin{eqnarray}
{\cal J}=-K_a(2P^{a}-{\cal
R}e^{a})=K_a{}^{\ast^{-1}}G^{a}\;,\nonumber
\end{eqnarray}
where $K_a{}^{\ast^{-1}}G^{a}$ (see also the equation (38) below) is
the current whose conservation is the starting point for the studies
referred in the introduction. The above analysis explicitly shows
that this current is the sum of the co-exact current $j_2$ given by
(32) and the co-closed current ${\cal J}_{2}=-K_a(P^{a}-{\cal
R}e^{a})$. It is perhaps, due to the lack of the relation (33), the
compositeness of the current $K_a{}^{\ast^{-1}}G^{a}$ have remained
unnoticed. As is shown in Appendix B, (33) directly follows from the
contracted Bianchi identities and (32).

\subsection{Currents for duals of Yano vector fields}

It is a well established fact that the Hodge map defines a general
one-to-one correspondence between the vector space of KY $p$-forms
and of the closed conformal KY $(n-p)$-forms \cite{ozumav1}. In
particular, for each KY $(n-1)$-form $\omega_{(n-1)}$ there exists a
uniquely determined closed conformal KY $1$-form $\tilde{Y}$ such
that
\begin{eqnarray}
\omega_{(n-1)}=^{\ast}\tilde{Y}\;,
\end{eqnarray}
where the conformal vector field $Y=Y^{a}X_a$, known as the {\it
Yano vector} \cite{McLenaghan}, is the metric dual of
$\tilde{Y}=Y_{a}e^{a}$ . Yano vectors are locally gradient fields
whose integral curves are pre-geodesics and they generate special
conformal transformations. For (34) we have
\begin{eqnarray}
{\cal J}_{2}=(-1)^{n-1}i_{X_{a}}(^{\ast}\tilde{Y}\wedge
P^{a})=i_{X_{a}}(\tilde{Y}\wedge^{\ast} P^{a})\;.
\end{eqnarray}
Since $i_{X^{a}}{}^{\ast} P^{a}=^{\ast}(P^{a}\wedge e_a)=0$ by the
Bianchi identity, we deduce from (35) that $Y_{a}^{\ast} P^{a}$ is
co-closed and equivalently, the $1$-form $Y_{a}P^{a}$ is closed.
Therefore, the vector field $Y_{a}\tilde{P}^{a}$ is locally a
gradient field (in fact, the equation (41) given below proves that
$Y_{a}\tilde{P}^{a}$ is globally a gradient field).

In order to write ${\cal J}_{1}$ in terms of KY $(n-1)$-forms we
first express $j_1$ and $j_2$ in terms of KY $(n-1)$-form
$^{\ast}\tilde{Y}$ and the Einstein $(n-1)$-forms $G^{a}$ as
follows;
\begin{eqnarray}
j_1&=&i_{X^a}i_{X^b}{^\ast}\tilde{Y}\wedge R_{ab}
=-Y_{a}G^{a}\;,\\
j_2&=&i_{X_a}{^\ast}\tilde{Y}\wedge P^{a} =Y^b{^\ast}e_{ab}\wedge
P^{a}=(-1)^{n}Y_{a}(G^{a}+{^\ast}P^{a})\;,
\end{eqnarray}
where
\begin{eqnarray}
G^{c}=R_{ab}\wedge^{\ast}(e^{cab})={\cal
R}^{\ast}e^{c}-2^{\ast}P^{c}\;.
\end{eqnarray}
So, by combining (36) and (37) as in (6), ${\cal J}_{1}$ is found to
be
\begin{eqnarray}
{\cal J}_{1}=-Y_{a}{^\ast}P^{a}\;.
\end{eqnarray}
By virtue of (24), this proves the co-exactness of
$Y_{a}{^\ast}P^{a}$:
\begin{eqnarray}
Y_{a}{^\ast}P^{a}=\frac{n-1}{n}\delta d{^\ast}\tilde{Y}\;.
\end{eqnarray}
or, equivalently, the exactness of $Y_{a}P^{a}$
\begin{eqnarray}
Y_{a}P^{a}=\frac{n-1}{n}d\delta\tilde{Y}\;.
\end{eqnarray}

Equations (32) and (41) reveal an important general role of Killing
and Yano vector fields that may be of significant in geometry as
well in physics. These vector fields are the ``integrating vector
fields" for Ricci $1$-forms, in the sense that if the underlying
space-time accepts such vector fields then their contractions with
Ricci $1$-forms as in (32) and (41) are, respectively, co-exact and
exact forms which are also specified in terms of the duals of these
vector fields. From (38) we also see that they enable us to make
decompositions of the corresponding contractions of the Einstein
$(n-1)$-forms which contain exact or co-exact parts. On the other
hand, for a KY $(n-2)$-form $\omega$ we can write
$\omega=^{\ast}\beta$, where $\beta$ is a closed conformal KY
$2$-form. In this case we have
\begin{eqnarray}
{\cal J}=i_{X_{a}}i_{X_{b}}(\omega\wedge
R^{ba})=\beta_{ba}{}^{\ast}R^{ba}\;,\nonumber
\end{eqnarray}
which exhibits the decomposition of $\beta_{ba}{}^{\ast} R^{ba}$ as
a sum of co-exact ${\cal J}_{1}$ and co-closed ${\cal J}_{2}$.

\section{Conserved Currents on Some Special Manifolds}

On a Ricci-flat (pseudo)Riemannian manifold for which $P^{a}=0$ and
hence ${\cal R}=0,\;{\cal J}_{2}$ is identically zero and ${\cal
J}_{1}$ is equal to $-j_1$. Therefore, for each KY $p$-form $\omega$
accepted by such a manifold we have
\begin{eqnarray}
(i_{X_a}i_{X_b}\omega)\wedge R^{ab}=\frac{p}{p+1}\delta
d\omega\;.\nonumber
\end{eqnarray}

Let us now consider a conformally flat (pseudo)Riemannian manifold
which is characterized by the vanishing of Weyl $2$-forms, or
equivalently, by
\begin{eqnarray}
R_{ab}=\frac{1}{n-2}[P_a\wedge e_b-P_b\wedge e_a+\frac{\cal
R}{n-1}e_a\wedge e_b]\;.\nonumber
\end{eqnarray}
This implies that $j_1$ linearly depends on $j_2$ and $j_3$ and
after a little computation we get
\begin{eqnarray}
{\cal
J}_{1}=(-1)^{p}\frac{n-2p}{n-2}j_2-\frac{p(p-1)}{(n-1)(n-2)}j_3\;.\nonumber
\end{eqnarray}
Thus, when $n$ is even and $p=n/2,\;j_3$ is co-exact. In general, in
view of (6) and (7) we can say that for $1<p<n-1,\;{\cal J}_{1}$ and
${\cal J}_{2}$ are linearly independent if and only if $j_2$ and
$j_3$ are.

\subsection{Constant-curvature space-times}

For an $n$-dimensional constant-curvature space-time \cite{Thirring}
characterized by $R_{ab}=ce_{a}\wedge e_{b}$ with $c$ constant, the
relations
\begin{eqnarray}
 P_{a}=c(n-1)e_{a}\;,\quad {\cal
R}=cn(n-1)\;.
\end{eqnarray}
directly follow from $R_{ab}=ce_{a}\wedge e_{b}$. In such a case we
have,
\begin{eqnarray}
{\cal J}_{1}&=&-c(i_{X^{a}}i_{X^{b}}\omega)\wedge
e_{ab}+(-1)^{p}c(n-1)i_{X_{a}}\omega\wedge e^{a}\nonumber\\
&=&-c p(n-p)\omega\;,\\
{\cal J}_{2}&=&(-1)^{p}c(n-1)i_{X_{a}}\omega\wedge
e^{a}+cn(n-1)\omega\nonumber\\
&=&c(n-1)(n-p)\omega\;,
\end{eqnarray}
where the scaling property has been used several times. Thus, the
currents become constant multiples of their defining KY-forms and
therefore they are linearly dependent. Moreover, their co-closedness
are nothing more than the co-closedness of KY-forms and hence give
nothing new. But the co-exactness of ${\cal J}_{1}$ has an important
implication.

By comparing (24) and (43) we obtain, for $p\neq 0$
\begin{eqnarray}
\delta d\omega=c(p+1)(n-p)\omega \;,\nonumber
\end{eqnarray}
and by recalling the fact that every KY-form is co-closed, we see
that every KY $p$-form is an eigenform of the Laplace-Beltrami
operator:
\begin{eqnarray}
\displaystyle{\not}d^2\omega=-c(p+1)(n-p)\omega.
\end{eqnarray}
This is nontrivial for $p\neq 0$ and $p\neq n$. The eigenvalues
depend on $n$, on the degree of the KY-form and on the constant $c$.
In particular, the sign of eigenvalues is the opposite of the sign
of $c$. Since the connection of a (pseudo)Riemannian geometry is
metric compatible, covariant derivatives commute with the Hodge map.
This implies that three main operators of section II A and therefore
$\displaystyle{\not}d^2$ also commute with the Hodge map. Equation
(45) shows that the Hodge dual of any KY $p$-form (which need not be
a KY $(n-p)$-form, but is certainly a closed conformal KY
$(n-p)$-form) is also an $(n-p)$-eigenform of
$\displaystyle{\not}d^2$ corresponding to the same eigenvalue. If we
restrict ourselves to $p$-eigenforms of the Laplace-Beltrami
operator we also get another series coming from the Hodge dual of a
KY $(n-p)$-form. So, for $\alpha=^{\ast}\beta$ where $\beta$ is a KY
$(n-p)$-form, we have
\begin{eqnarray}
\displaystyle{\not}d^2\alpha=-cp(n-p+1)\alpha,\nonumber
\end{eqnarray}
where $\alpha$ is a closed conformal KY $p$-form. As a result, each
eigenvalue corresponding to a KY $p$-form is degenerate with
multiplicity (counted in the space of $p$-forms) which is not lower
than the number of linearly independent KY $p$-forms. An exceptional
case may occur when $n$ is even and $p=n/2$. In such a case, the
mentioned lower bound may decrease depending on the existence of
dual and anti-self dual KY $p$-forms. By now, it is well-established
fact that the number of linearly independent KY $p$-forms is bounded
from above, for any dimension and signature, by the binomial number
\begin{eqnarray}
C(n+1, p+1)=\frac{(n+1)!}{(p+1)!(n-p)!}\;,
\end{eqnarray}
and these upper bounds are attained on the constant curvature
manifolds \cite{Kastor,ozumav1}.

In fact, on a constant curvature Riemannian manifold, such as the
standard $n$-sphere, the spectrum of $\displaystyle{\not}d^2$ on
$p$-forms is well known \cite{Semmelman1,Semmelman2}: it consists of
two series related to each other by the interchange
$p\leftrightarrow (n-p)$ in the eigenvalues and both depend (apart
from $n$ and $p$) on a nonnegative integer $k=0,1,\dots$. Eigenforms
corresponding to the minimal values of these series (for which
$k=0$) turn out to be KY $p$-forms and the Hodge duals of KY
$(n-p)$-forms. What we have found by analyzing one of the currents
above correspond to minimal parts of the spectrum for the case of a
constant curvature space-time which need not necessarily be
Riemannian.

\subsection{Einstein Manifolds}

Let us now consider an $n$-dimensional Einstein manifold
characterized by $P^{a}=ke^{a}$ with constant $k={\cal R}/n$. Since
in three dimensions any Einstein space is necessarily of constant
curvature here we suppose $n\geq 4$ and $R_{ab}\neq ce_a\wedge e_b$
to reach nontrivial statements. In this case we have
\begin{eqnarray}
{\cal J}_{1}&=&-i_{X^{a}}i_{X^{b}}\omega\wedge R_{ab}-k p \omega\;,\nonumber\\
{\cal J}_{2}&=&k(n-p)\omega\;.\nonumber
\end{eqnarray}
Obviously, the conservation of ${\cal J}_{2}$ becomes trivial but,
the co-exactness of ${\cal J}_{1}$ implies
\begin{eqnarray}
j_{1}&=&i_{X^{a}}i_{X^{b}}\omega\wedge R_{ab}\;,\nonumber\\
&=&kp\omega-\frac{p}{p+1}\delta d\omega\;.\nonumber
\end{eqnarray}

In particular, for any KY $2$-form $\omega=(1/2)\omega_{ab}e^{ab}$
of an Einstein manifold the ``contracted'' curvature $2$-form
$\omega_{ba} R^{ab}$ can be written as
\begin{eqnarray}
\omega_{ba}R^{ab}=2k\omega-\frac{2}{3}\delta d\omega\;.
\end{eqnarray}
Note that this is trivial for constant curvature spaces. This
relation shows that for each KY $2$-form accepted by an Einstein
space the contraction of the curvature $2$-forms as in equation (47)
is decomposed into co-closed and co-exact parts that are determined
up to well-defined constants, by these KY $2$-forms. On the other
hand, the equations (32) and (40) (or (41) in the case of Einstein
manifolds imply that $\tilde{K}$ and $\tilde{Y}$ (and hence
$^\ast\tilde{K}$ and ${}^\ast\tilde{Y}$) are eigenforms of the
Laplace-Beltrami operator:
\begin{eqnarray}
\displaystyle{\not}d^2\tilde{K}=
-2k\tilde{K}\;,\quad\displaystyle{\not}d^2\tilde{Y}=k\frac{n}{n-1}\tilde{Y}\;.
\end{eqnarray}
Note that while in the first case the eigenvalues have the opposite
sign with $k$, in the second case they have the same signs. The
lower bounds for their degeneracy can be determined from (46) as we
did in the previous subsection.

\section{Summary and Concluding Remarks}

In this study we have shown that two linearly independent
uncomposite currents for each KY-form can be constructed using the
curvature forms of the underlying space-time. The current suggested
in the literature (see (1)) is a special linear combination of the
currents claimed here. Moreover, while the current (1) identically
vanishes for all KY $(n-1)$-forms, in our case the currents ${\cal
J}_{1}$ and ${\cal J}_{2}$ are non-vanishing but become linearly
dependent. What is more, the current ${\cal J}_{1}$ is shown to be
co-exact from which several
geometric facts follow. Some of these can be summarized as follows.\\
(i) The scalar curvature ${\cal R}$ of any (pseudo)Riemannian
manifold is a flow invariant for
all of its Killing vector fields; $\nabla_{K}{\cal R}=0$. \\
(ii) The conserved currents directly provide decompositions for
index-saturated differential forms constructed by contracting the
components of KY $p$-forms  with the Ricci $1$-forms, curvature
$2$-forms and with the Einstein $(n-1)$-forms (for the values
$p=1,2$ and $p=n-1$). These contractions are solely expressible in
terms of KY-forms themselves and their (co)derivatives. \\
(iii) On an Einstein manifold, duals of all Killing and Yano vector
fields are eigenforms of the Laplace-Beltrami operator such that the
eigenvalues have multiplicities with well defined lower bounds.\\
(iv) Generalizations of some well-known facts related to the
spectrum of Laplace-Beltrami operator on a constant curvature
Riemannian manifold also follow directly from the properties of
conserved currents. More precisely, we have shown that in a constant
curvature (pseudo)Riemannian manifold all KY-forms and their Hodge
duals are eigenforms of the Laplace-Beltrami operator.

The whole attention in this study has inevitably been focused on the
basic gravitational currents and their properties. The study of
related conserved charges in the framework of this paper remains
almost untouched, but deserves to be the subject of a separate
study. The basic gravitational currents presented in this paper may
shed light on some of the problems related to interpretation of
generalized charges.

\begin{acknowledgments}
This work was supported in part by the Scientific and Technical
Research Council of Turkey (T\"{U}B\.{I}TAK).
\end{acknowledgments}

\begin{appendix}

\section{Contracted Bianchi Identities}

In (pseudo)Riemannian geometry, the identities
\begin{eqnarray}
R^{a}_{\;b}\wedge e^{b}&=&0\;,\\
DR^{ab}\equiv dR^{ab}+\omega^{a}_{\;c}\wedge
R^{cb}+\omega^{b}_{\;c}\wedge R^{ac}&=&0\;,
\end{eqnarray}
are known as the first and the second Bianchi identities,
respectively. Here $D$ denotes the covariant exterior derivative.
The following useful identities can easily be verified by taking
successive contractions of (A1) \cite{Benn-Tucker};
\begin{eqnarray}
i_{X_a}i_{X_b}R_{cd}&=&i_{X_c}i_{X_d}R_{ab}\;,\quad
i_{X_a}P_b=i_{X_b}P_a\;,\\\
i_{X_a}R_{bc}+c.p.&=&0\;,\qquad \qquad \;\;P_b\wedge e^{b}=0\;.
\end{eqnarray}
Here c.p. stands for the cyclic permutations. There are also
additional contracted Bianchi identities that can be obtained from
(A2). But, these are scattered in the literature and since they are
heavily used in our analysis, it will be convenient to give some
details of their derivations.

It turns out that the Bianchi identities implied by (A2) can most
concisely be expressed in terms of $2$-form $T_{c}{}^{ab}$ and its
contraction defined by
\begin{eqnarray}
T_{c}{}^{ab}&=&\nabla_{X_c}R^{ab}+{\omega^{a}}_d(X_c)R^{db}+{\omega^{b}}_d(X_c)R^{ad}\;,\\
S_{c}{}^{b}&=&i_{X_a}T_{c}{}^{ab}\;.
\end{eqnarray}
By definition, $T_{c}{}^{ab}$ is anti-symmetric in the last two
indices
\begin{eqnarray}
T_{c}{}^{ab}=-T_{c}{}^{ba}\;.
\end{eqnarray}
Since
\begin{eqnarray}
i_{X_a}T_{c}{}^{ab}=\nabla_{X_c}(i_{X_a}R^{ab})+{\omega^{b}}_d(X_c)P^{d}+{\omega^{a}}_d(X_c)i_{X_a}R^{db}-i_{\nabla_{X_c}X_a}R^{ab}\;,\nonumber
\end{eqnarray}
and as the last two terms cancel, we obtain
\begin{eqnarray}
S_{c}{}^{b}=\nabla_{X_c}P^{b}+{\omega^{b}}_d(X_c)P^{d}
\end{eqnarray}
By contracting this relation with $X_b$ we find
\begin{eqnarray}
i_{X_b}S^{cb}=\nabla_{X^c}\cal{R}.
\end{eqnarray}
 As an aside, we should note that (A2) can be rewritten as
$e^{c}\wedge {T_c}^{ab}=DR^{ab}=0$ and (A8) implies that
$e^{c}\wedge {S_c}^{a}=DP^{a}$.

We are now ready to take the interior derivative of (A2) and write
\begin{eqnarray}
T^{cab}&=&di_{X^c}R^{ab}+{\omega^{a}}_d\wedge
i_{X^c}R^{db}+{\omega^{b}}_d\wedge i_{X^c}R^{ad}+\omega^{cd}\wedge
i_{X_d}R^{ab}\;,
\end{eqnarray}
where we have used (A3). This can be rewritten in a more compact
form as
\begin{eqnarray}
T^{cab}=Di_{X^{c}}R^{ab}\;.
\end{eqnarray}
In view of the first relation of (A4) and (A11) easily follows that
\begin{eqnarray}
T^{cab}+T^{abc}+T^{bca}=0.
\end{eqnarray}
Two successive contractions of this relation yield
\begin{eqnarray}
i_{X_a}T^{abc}-S^{bc}+S^{cb}&=&0\;,\\
i_{X_c}S^{cb}-i_{X_c}S^{bc}+i_{X_c}i_{X_a}T^{abc}&=&0\;.
\end{eqnarray}
Noting that the third term of (A14) is equal to the first term, by
virtue of (A9) we obtain
\begin{eqnarray}
i_{X_c}S^{cb}=\frac{1}{2}\nabla_{X^{b}}\cal{R}.
\end{eqnarray}
From (A9) and (A15) we observe an interesting property of $S^{cb}$:
its contraction with respect to second index is twice the
contraction with respect to first index.

As a result; (A10) (or, equivalently (A11)), (A12), (A13) and (A15)
are additional contracted Bianchi identities resulting from
contraction of the second Bianchi identity given by (A2). In
particular, the last one will play a prominent role in the next
appendix.

\section{An Alternative Proof of $\nabla_{K}{\cal{R}}=0$}

Our alternative proof of (33) proceeds as follows. From (A8) we
obtain
\begin{eqnarray}
i_{X_c}S^{cb}=i_{X_c}\nabla_{X^{c}}P^{b}+{\omega^{b}}_{d}(X_c)i_{X^{c}}P^{d}\;,\nonumber
\end{eqnarray}
and by using this in (A15), the covariant derivative of ${\cal R}$
with respect to an arbitrary Killing vector field $K$ is found to be
\begin{eqnarray}
\nabla_{K}{\cal
R}=2K_{b}i_{X_c}S^{cb}=2i_{X^{c}}[\nabla_{X_c}(K_bP^{b})-P^{b}\nabla_{X_c}K_b]+
2K_b\;{\omega^{b}}_{d}(X_c)i_{X^{c}}P^{d}.
\end{eqnarray}
In view of (32) and $\delta^2=0$, the first term of the second
equality vanishes and (B1) reduces to
\begin{eqnarray}
\nabla_{K}{\cal R}=-2(i_{X^{c}}P^{b})i_{X_b}\nabla_{X_c}\tilde{K}\;.
\end{eqnarray}
In writing this relation we renamed the indices and made use of
\begin{eqnarray}
i_{X_b}\nabla_{X_c}\tilde{K}=\nabla_{X_c}K_b+K_d\;{\omega_b}^{d}(X_c)\nonumber.
\end{eqnarray}
By the second identity of (A3) and since the symmetrized covariant
derivative of any KY-form is zero we obtain $\nabla_{K}{\cal{R}}=0$
from (B2).

\end{appendix}

%\references%

\end{document}